Dhairya Patel[a]* and Shaifali P. Malukani[b]

[a]*Department of Computer Engineering, Dharmsinh Desai University, Nadiad, India;*
[b]*Department of Computer Engineering, Dharmsinh Desai University, Nadiad, India*

*\*Dhairya Patel – [dpvp1403@gmail.com](mailto:dpvp1403@gmail.com)*

# Navigating Fog Federation: Classifying Current Research and Identifying Challenges.

Fog computing has gained significant attention for its potential to enhance resource management and service delivery by bringing computation closer to the network edge. While numerous surveys have explored various aspects of fog computing, there is a distinct gap in the literature when it comes to fog federation—a crucial extension that enables collaboration and resource sharing across multiple fog environments, enhancing scalability, service availability, and resource optimization. This paper provides a comprehensive survey of the existing work on fog federation, classifying the contributions from its inception to the present. We analyse the various approaches, architectures, and methodologies proposed for fog federation and identify the primary challenges addressed in this field. In addition, we explore the simulation tools and platforms utilized in evaluating fog federation systems. Our survey uniquely contributes to the literature by addressing the specific topic of fog federation, offering insights into the current state of the art and highlighting open research gaps and future directions.

Keywords: fog computing; fog federation; iot;

Subject classification codes: 68Q85

**I) Introduction**

The rapid growth of the Internet of Things (IoT) has led to the emergence of fog computing as a critical paradigm for managing distributed resources and processing data closer to where it is generated [1]. By extending cloud capabilities to the edge of the network, fog computing reduces latency, enhances real-time processing, and improves bandwidth usage. However, as the number of fog users increases, isolated fog environments may struggle to meet the growing demand for resources and services [2].

Fog federation addresses these limitations by enabling multiple fog environments to collaborate and share resources. Figure 1.1 illustrates the basic IoT- Cloud architecture, whereas Figure 1.2 depicts the Fog Federation architecture, where Fog Nodes collaborate across domains to process data locally before offloading to the cloud for further analysis.

This approach enhances scalability, optimizes resource utilization, and ensures better service availability across the federation. By integrating geographically distributed fog nodes, a federation can balance load, manage failures, and improve overall performance, making it a critical extension of the fog computing model.

While there has been extensive research on various aspects of fog computing [3], fog federation remains relatively unexplored. Authors in [4] explored the underlying architectures of fog systems, focusing on key features such as scalability, security, and privacy, as well as the Quality of Service (QoS) parameters that ensure optimal performance. Article [5] identifies and consolidates the technical, non-functional, and economic obstacles encountered in the adoption of fog computing. Meanwhile, Article [6] explores theintegration of databases within edge and fog environments, discussing the benefits and challenges facedby developers. The server placement strategies across edge, fog, and cloudlet environments are thoroughly analysed in Article [7], which also highlights strengths, weaknesses, and future research opportunities. Article [8] focuses on vehicular fog computing (VFC) architectures, emphasizing issues of trust, security, and real-time big data analytics. Article [9] reviews various approaches to QoS and energy consumption in fog computing, while Article [10] delves into the architecture of fog computing for IoT (FC-IoT), addressing communication, security, and decision-making aspects. Article [11] systematically reviews fog computing applications in the healthcare sector. Despite its potential to revolutionize distributed computing, there is a distinct lack of comprehensive surveys dedicated to the field of fog federation—a concept that involves the coordination and collaboration of multiple fog networks to optimize resource use and service delivery. This research addresses the gap by exploring the challenges of fog federation, offering a detailed classification of existing studies, and proposing future research directions. To the best of our knowledge, it is the first survey dedicated solely to fog federation. We classify the existing research in the field of fog federation, analyse the architectural frameworks, identify key challenges, and explore the tools used for simulation and evaluation. In doing so, we offer insights into the state of fog federation research and highlight opportunities for future development in this promising area.

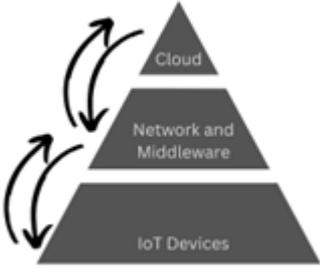
Figure 1.1 Traditional Cloud-IoT architecture

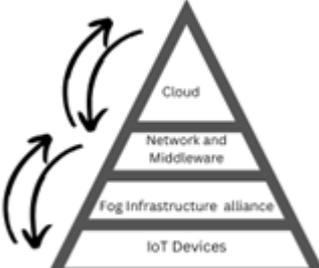
Figure 1.2 Fog Federation Supported Cloud-IoT architecture

Table 1. Cloud-IoT architecture

This paper is organized as follows: Section 2 discusses the architecture of fog federation. Section 3 outlines the research methodology. Section 4 presents a classification of existing studies. Section 5 reviews the simulation tools used in evaluating fog federation systems. Section 6 highlights current challenges and suggests future research directions. Finally, Section 7 concludes the paper.

**II) Architecture of Fog Federation System**

The fog federation system architecture, as illustrated in Figure 2, is organized into three hierarchical layers: End Devices, a Fog Node Layer (divided into multiple Fog Domains), and a Cloud Layer. This structure promotes resource sharing, computational load balancing, and low-latency communication. The key feature of this architecture is the fog federation, which enables dynamic collaboration between fog domains to optimize resource usage and ensure seamless service delivery.

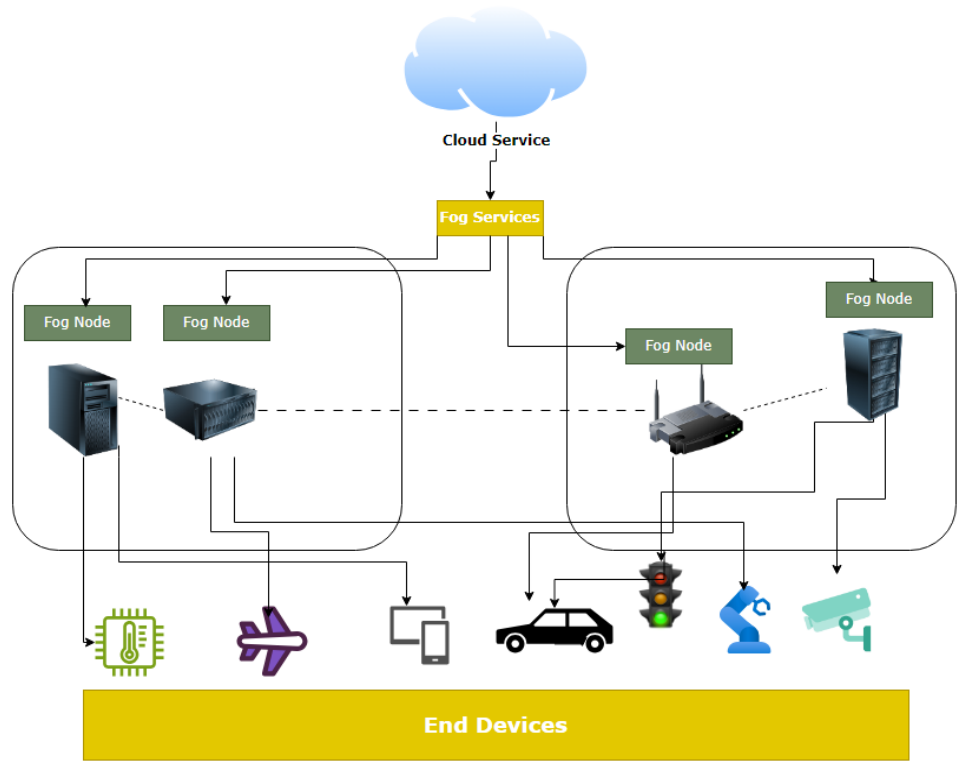

Figure 2: Fog Federation Architecture

*End Devices Layer*

The End Devices Layer consists of resource-constrained devices, such as sensors and smart appliances, which generate real-time data but lack the computational capacity to process it independently. These devices rely on the fog federation for data processing and decision-making. The fog nodes receive data from these end devices, ensuring rapid responses to immediate requirements like environmental monitoring or smart system control.

*Fog Node Layer*

The Fog Node Layer consists of multiple domains with fog nodes handling local processing, storage, and decision-making. These nodes manage real-time analytics, traffic routing, and emergency responses, while the fog federation enables inter-domain collaboration to optimize resource use, ensuring scalability and adaptability.

*Cloud Layer*

The Cloud Layer handles large-scale storage, complex computations, and long-term data retention. When fog nodes reach their processing limits, tasks are offloaded to the cloud for further analysis. It also ensures system backup and disaster recovery.

*Operational Flow*

End devices send data to nearby fog nodes for processing. The fog federation enables dynamic resource sharing between domains for optimal performance. For complex tasks, data is offloaded to the cloud. This layered system reduces latency, improves real-time responsiveness, and ensures efficient service delivery.

**III) Research Methodology**

This review employs a systematic approach to assess the current research on fog federation. It identifies key challenges, evaluates simulation tools, uncovers research gaps, and explores future directions. With fog federation research in its early stages, this survey presents a thorough collection of all key papers since its inception.

*Review Methodology*

We conducted a systematic search across academic databases like IEEE Xplore, ACM Digital Library, SpringerLink, and Google Scholar. Key terms such as "fog federation," "fog computing," and "simulation tools" guided our search. We focused on peer-reviewed articles, conference papers, and technical reports published since the inception of fog computing i.e. between 2012 and 2024, excluding non-English articles and those lacking detail.

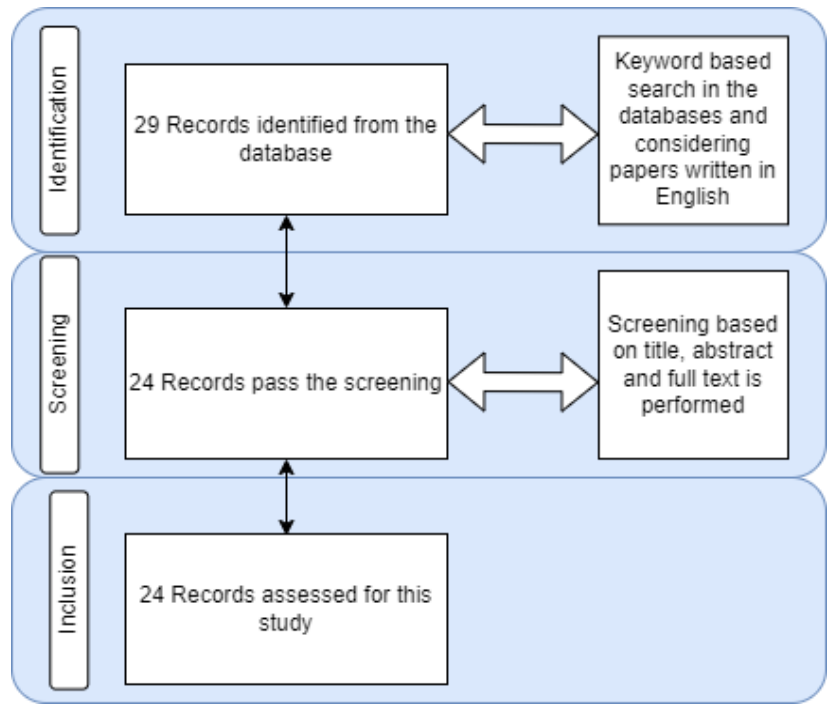

Figure 3: Review Methodology

Titles and abstracts were screened, followed by a detailed review of full texts to confirm relevance. We extracted information on research objectives, methodologies, challenges, tools, gaps, and future directions. Figure:3 illustrates the different stages of the review methodology, providing a visual representation of the process, from initial search to data extraction. Research studies included in this survey were sourced from reputable publications such as ACM, IEEE, Springer, Elsevier, and various other scholarly journals. Figure: 4 illustrates the number of research articles associated with each publication.

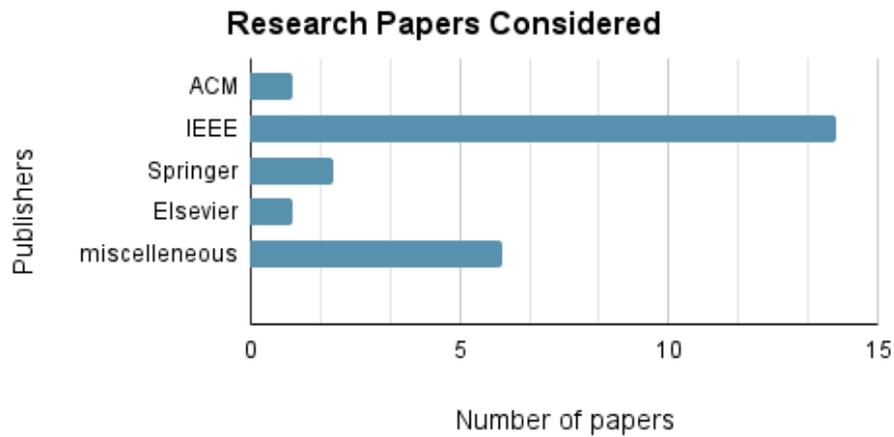

Figure 4: Reviewed papers

Research Questions: The review aims to address critical questions regarding the challenges, evaluation tools, research gaps, and future directions in fog federation. These questions seek to advance understanding of the field and highlight areas not previously explored in detail.

- What are the primary challenges currently being addressed in fog federation research?
- What simulation tools are currently available for evaluating fog federation systems?
- What are the major research gaps identified in the current literature on fog federation?
- What future directions are proposed to overcome the existing challenges in fog federation?

**IV) Classification of Existing Work in Fog Federation**

As shown in figure 5 Research in fog federation can be classified into three primary areas: federation formation, federation management, and security. Federation formation focuses on enabling collaboration between fog nodes, while federation management addresses key operational aspects such as task scheduling, application placement, and offloading strategies. Security remains a significant area of focus, encompassing key exchange, authentication, access control, and confidentiality to ensure secure and efficient system functionality.

*Fog Federation Formation*

Fog Federation Formation Techniques refer to the methods and strategies used to organize and establish a network of fog nodes into a cohesive and efficient federated system. These techniques involve determining how to group and connect fog nodes across different geographical areas to achieve optimal resource sharing, enhance performance, and ensure effective communication and collaboration among the nodes.

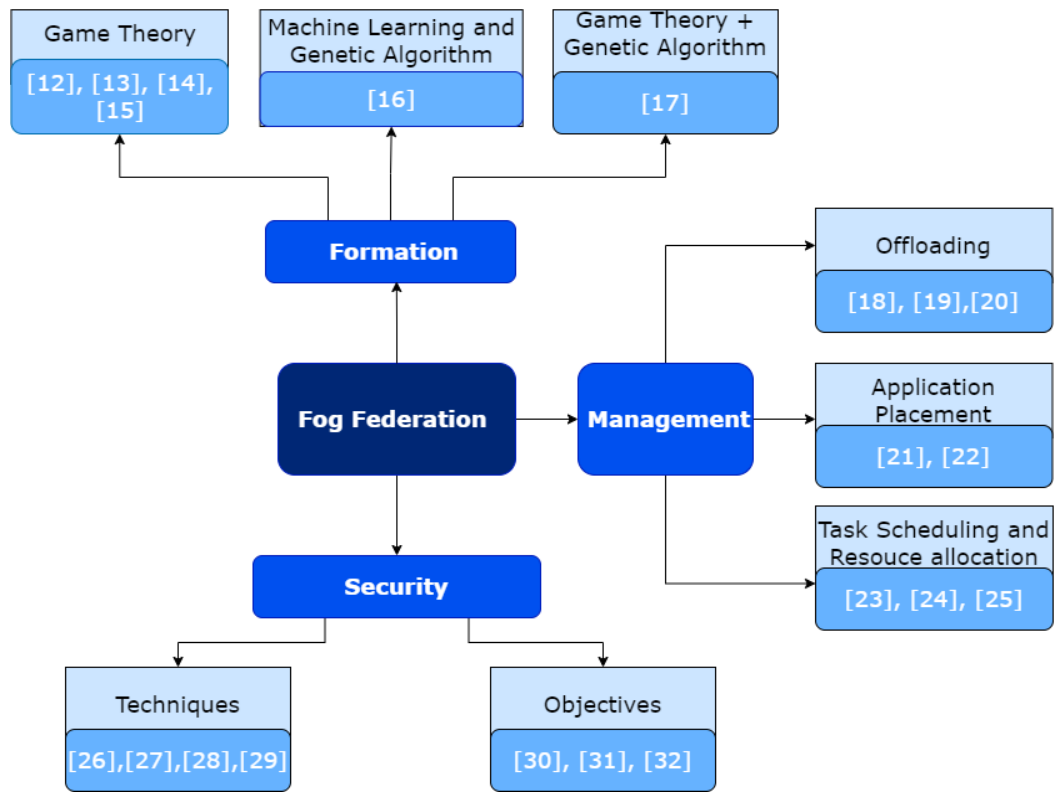

Figure 5: Fog Federation Research Classification

Fog federation formation methods incorporates both game-theoretic and genetic algorithm-based approaches to enhance fog federation systems, addressing challenges such as resource allocation, stability, Quality of Service (QoS), and overall system performance. Game-theoretic approaches play a significant role in ensuring efficient resource sharing, stable federation structures, and optimizing outcomes for participants within fog federations. For instance, work done in [12], coalition formation games are used to enable cooperative workload and resource sharing among Fog Infrastructure Providers (FIPs). By employing a coalition value function, the model determines the desirability and stability of coalitions, optimizing resource allocation structures and improving QoS for IoT applications under varying workloads. Future work in this area involves refining the coalition value function to consider revenue losses from insufficient resources and further validating the model on real testbeds.

The work in [13] introduces a game-theoretic approach based on Hedonic Coalition Formation to create adaptive fog federations for the Internet of Vehicles (IoV). The architecture is decentralized and dynamically adjusted through a Merge & Split technique to maintain QoS despite the mobile nature of vehicles. Experimental results using SUMO simulator data demonstrate that this approach enhances user satisfaction, stability, and execution efficiency compared to existing models. The work in [14] research introduces a federated learning architecture enhanced by fog federations for the Internet of Vehicles (IoV), leveraging Hedonic coalition games to ensure stable federations and improve model accuracy and service quality. Simulations for traffic sign detection show that this approach outperforms existing methods. Research [15] tackles instability in fog federations by introducing a decentralized algorithm based on the Replicator Dynamics model, enhancing federation stability. Fog federations offer advantages such as improved resource sharing, scalability, and load balancing, but instability can degrade performance and service availability. This study is unique in addressing federation stability through an evolutionary game-theoretical model and a decentralized algorithm.

In contrast, in some of the approaches genetic algorithms and machine learning are utilized in fog federations to optimize resource allocation, enhance system performance, and improve computational efficiency. The work in [16] presents a novel federated fog architecture that employs Genetic Algorithms and AI-based models to improve throughput and reduce response time. This architecture addresses QoS issues in IoT by adapting federation formation to ensure efficient load distribution and minimize delays for real-time services. Evaluations with real data show significant improvements in performance. Future work will explore advanced machine learning techniques like Reinforcement Learning and Deep Learning to further enhance the system.

Work in [17] incorporates a dual approach, combining genetic algorithms with game-theoretic models for effective resource management and enhanced computational efficiency within fog federations. This method for forming fog federations that addresses computational complexity, user data privacy, and decentralized management. Utilizing federated learning, the approach predicts Quality of Service (QoS) while preserving data privacy. It features a weighted algorithm for optimal client selection, a genetic algorithm for maximizing profit and QoS, and an evolutionary game-theoretic technique for federation stability. The proposed method enhances QoS and stability compared to existing approaches. Future challenges include managing federations without a broker, which may present unforeseen operational issues despite eliminating broker-related costs.

*Fog federation management*

This section focuses on the management aspects of fog federation, including task offloading, application distribution, and task scheduling. Understanding these aspects is crucial for optimizing resource utilization, ensuring scalability, and maintaining high Quality of Service (QoS) within federated fog environments. Effective management of these elements is essential for enhancing overall system performance and efficiency in fog computing. Figure 5 shows the studies focusing on these aspects of fog federation.

*Offloading*

Offloading refers to the process of transferring tasks or computational workloads from a primary fog node to another fog node. This technique is used to manage and optimize resources by distributing tasks to more capable or strategically located systems, thereby improving performance, reducing latency, and alleviating the computational burden on the originating fog node.

In research [18], the authors explore methods to maximize profit through efficient fog node utilization. This study introduces a policy for offloading delay-sensitive tasks in a fog network while simultaneously maximizing revenue for service providers. The proposed two-stage policy involves initially allocating computational resources to nearby end-users and subsequently using a Fog Manager to exploit any unused resources for additional revenue. Future work aims to incorporate considerations for inter-fog transmission delays and pricing agreements among service providers to enhance the handling of deadline-aware tasks. research [19] proposes a federated load-sharing model for the Internet of Vehicles (IoV). This micro-level fog deployment model aims to manage heavy workloads through balanced resource sharing. It has been evaluated against traditional random walk and neighbouring fog selection algorithms, showing significant improvements in reduced delay, queue length, and workload balance across the fog federation.

In research [20], Fog Federation is employed to offload video streaming tasks, achieving a latency reduction of over 52% compared to traditional CDNs. The study details a federated fog network (F-FDN) where video requests are managed by the Central Cloud and processed by fog nodes, utilizing cached content and neighbouring nodes to optimize delivery. Future research will explore a multi-tiered structure within the F-FDN framework.

*Application Placement*

Application placement refers to the strategic positioning of applications or their components across different computing resources or nodes within a fog federation network. Application placement is crucial as it optimizes resource utilization, improves performance, and ensures efficient load balancing and scalability in distributed computing environments. Research [21] and Research [22] both address this aspect of fog federation, focusing on optimizing resource utilization and enhancing system performance.

Research [21] demonstrates the use of Kubernetes cluster federation in Fog Computing. The KubeFed tool facilitates the management of multiple Kubernetes clusters as a unified entity, enhancing the elasticity and resilience of fog computing systems. The study describes a two-phase workload placement mechanism that effectively distributes microservices across federated infrastructure.

Another practical application of Fog Federation is discussed in research [22], which addresses the need for optimal resource allocation within Industry 4.0 contexts. The paper introduces a load-balancing approach that considers both the software architecture of applications and their deadlines. Future goals include more granular partitioning within each microservice to further optimize resource allocation.

*Task scheduling and resource allocation*

Task scheduling involves organizing and timing the execution of tasks to optimize performance and meet deadlines, while resource allocation is the process of assigning available resources to tasks to ensure efficient utilization and performance. Task scheduling and resource allocation are crucial aspects of fog federation because they ensure that tasks are efficiently distributed and executed across multiple fog nodes, optimizing resource use, reducing latency, and enhancing overall system performance and scalability.

In research [23], an effective method for ordering task accomplishments based on their assigned rank has been presented, demonstrating superior efficiency compared to the traditional Round-Robin scheduling, which cyclically assigns tasks in succession. The study suggests that further improvements could be made by incorporating additional metrics crucial to federation management, such as energy consumption and system resilience. Work in [24] offers a notable approach to task placement strategies by proposing a Deadline-Aware Dynamic Task Placement (DDTP) method. This approach utilizes a Master-Slave architecture where a Fog Controller operates as the master, and Fog nodes function as slaves. The DDTP strategy prioritizes tasks based on their deadlines and employs a dispatch-constrained policy to reassign failed tasks to available fog nodes, thus optimizing task management within the federation. Work in [25] addresses the challenge of pricing within fog federations using the Stackelberg model. Fog devices bundle resources into Computing Resource Blocks (CRBs) and set prices, while IoT devices decide on CRB purchases. The study examines how IoT devices can minimize costs and how CRB pricing can maximize revenue for fog federations.

*Fog federation security*

The security of fog federation systems is paramount, with research dedicated to mitigating potential vulnerabilities through secure key exchange, robust authentication mechanisms, and strict access control protocols. Additionally, confidentiality and data availability are prioritized to safeguard sensitive information and maintain system integrity, ensuring secure communication and reliable operation across the federation. Table 1 presents an overview of the current research efforts in security aspect of fog federation.

|                   | [27] | [30] | [28] | [29] | [26] | [32] | [31] |
|-------------------|------|------|------|------|------|------|------|
| Key exchange      | Y    |      | Y    |      | Y    |      |      |
| Authentication    |      |      | Y    | Y    |      |      |      |
| Access Control    |      | Y    |      |      |      |      | Y    |
| Confidentiality   |      | Y    |      |      |      | Y    | Y    |
| Data Availability |      | Y    |      |      |      | Y    | Y    |

Table 2. Fog Federation Security

Work in [26] has introduced the Diffie–Hellman protocol based enhanced key exchange scheme tailored for Fog Federation. This scheme, validated using the Automated Validation of Internet Security Protocols and Applications (AVISPA) tool, demonstrates improved security against active and passive attacks. Additionally, it maintains low communication and moderate computational costs compared to existing protocols.

Expanding on this, [27] proposes a lightweight key exchange solution designed to reduce computational and communication overheads. This scheme improves efficiency by minimizing the computational load and communication requirements, making it more suitable for fog computing environments. Evaluated using AVISPA and simulated with the NS3 tool, the lightweight scheme shows significant reductions in both overheads while maintaining robust security.

In the Internet of Vehicles (IoV), secure authentication is crucial for ensuring data integrity and preventing unauthorized access. The Secure Authentication based on Fog-Cloud (SAIFC) scheme [28] addresses these needs by utilizing HTTP-based authentication and cookies. The scheme's effectiveness is confirmed through both formal analyses using AVISPA and simulations with the NS3 tool. The results indicate that the SAIFC scheme successfully reduces packet loss and improves throughput compared to existing solutions.

Furthermore, [29] proposes a universal fog proxy that supports third-party authentication across multiple fog providers using protocols like OpenID Connect (OIDC), 802.1x, and PANA. This proxy simplifies user access with a single account and effectively handles authentication requests.

Ensuring secure data access and managing rogue nodes are critical for protecting data in fog computing. [30] introduces a scheme combining Ciphertext-Policy Attribute-Based Encryption (CP-ABE) with cryptographic hashing, which categorizes fog nodes and divides sensitive files to mitigate the impact of rogue nodes. This approach ensures robust data protection and low communication overhead. The scheme [31] combines CP-ABE with a private blockchain to detect and prevent rogue fog nodes, minimizing communication overhead between fog nodes and the cloud server. It enables secure, distributed authorization in a trust-less environment, ensuring data confidentiality and availability.

The work in [32] presents a Cognitive Fog (CF) model designed to enhance security in smart healthcare services within IoT systems. It incorporates an ensemble classifier combining K-NN, DBSCAN, and Decision Tree (DT) techniques to effectively detect and classify attacks. Additionally, the paper implements Planned and Ad-hoc

Federations to improve communication, threat awareness, and process recovery, thereby ensuring data confidentiality and availability.

**V) Simulation Tools for Fog Federation**

In the realm of fog computing, effective simulation tools are crucial for modeling, analyzing, and optimizing the federation of distributed computational resources. This section reviews several prominent simulation tools and their relevance to fog computing federation.

*Ns3(Network Simulator 3)*

Ns3 is a discrete-event network simulator that offers a robust environment for simulating network behaviours and protocols. Its detailed event scheduling and network modeling capabilities are particularly relevant for fog computing environments.: Ns3 enables researchers to simulate the intricate network interactions between fog nodes, including communication delays, packet losses, and network topology changes. These simulations are crucial for understanding how fog nodes interact within a federated system, optimizing network performance, and ensuring efficient data routing and resource sharing across the fog infrastructure.

*AVISPA (Automated Validation of Internet Security Protocols and Applications)*

AVISPA is an automated tool designed for the validation of security protocols with minimal user intervention. It plays a critical role in ensuring the security and integrity of communication protocols.

In fog computing, security is a major concern due to the distributed nature of resources across various administrative domains. AVISPA can be used to validate security protocols employed in fog federations, such as those securing inter-node communication and data exchanges. By ensuring that security measures are robust and effective, AVISPA helps safeguard the integrity and confidentiality of federated fog computing environments.

## *Blockchain-based Brokerage Platform for Fog Computing Resource Federation[33]*

This tool leverages blockchain technology to facilitate decentralized resource management and federation across multiple administrative domains.

The Blockchain-based Brokerage Platform addresses the challenge of resource management in fog computing by enabling dynamic, transparent, and secure resource sharing across different domains. Through the use of smart contracts, it ensures that resource transactions are automated and reliable, which is essential for creating efficient and scalable federated fog computing systems. This decentralized approach mitigates the limitations of traditional centralized resource management and enhances cross-domain collaboration.

## *XFogSim[34]*

XFogSim is a specialized simulator designed to handle the complexities of fog computing environments, including latency, network delays, and packet error rates. XFogSim's features are highly pertinent to fog federation as it allows for the simulation of various performance metrics crucial to fog computing, such as latency and resource availability. Its ability to support locality-aware distributed broker node management makes it particularly useful for optimizing resource allocation and minimizing latency in federated fog environments. By providing insights into the trade-offs between cost, performance, and availability, XFogSim helps in designing and managing efficient fog federations.

### *iFogSim[35]*

iFogSim is a comprehensive simulation toolkit specifically designed for fog computing environments. It allows for the modeling of various aspects of fog computing, including resource allocation, application deployment, and the interaction between fog and cloud layers. iFogSim is particularly relevant for simulating and evaluating the federation of fog computing resources. It provides detailed modeling capabilities for different types of fog nodes, including sensors, edge devices, and cloud servers. The toolkit supports the simulation of resource provisioning, load balancing, and application placement across fog and cloud resources, enabling researchers to study the efficiency and effectiveness of resource federation strategies. iFogSim's ability to model complex scenarios involving multiple fog nodes and their interactions makes it a powerful tool for optimizing federated fog systems.

### *The Cloud Visitation Platform (CVP)[36]*

The Cloud Visitation Platform (CVP) enhances cloud virtualization by integrating hardware awareness into application design. Unlike traditional models that conceal hardware details, the CVP's layered architecture features distinct APIs and interfaces for managing hardware access, ensuring fair resource allocation and resilience during failures. It supports both pull-based and push-based deployment strategies, making it applicable for single-cloud providers and federated cloud environments. This approach enables applications to optimize resource utilization based on local conditions, improving overall performance and efficiency. Adopting the CVP paradigm allows cloud providers to bridge the gap between high-end federations and edge computing, promoting more adaptable cloud applications.

**VI) Challenges in Fog Federation**

Fog federation can offer significant benefits in terms of scalability, flexibility, and resource optimization, it also comes with several challenges:

(1) Interoperability: Different fog environments may use varied hardware, software, and communication protocols. Ensuring seamless interoperability between these diverse systems can be complex.

(2) Security: Federating fog environments introduces additional security concerns, including data privacy, access control, and secure communication. Ensuring that security measures are consistent across all federated nodes is crucial.

(3) Data Management: Coordinating data across multiple fog nodes involves managing data consistency, synchronization, and storage. Handling data locality and ensuring efficient data transfer and processing can be challenging.

(4) Resource Allocation: Efficiently managing and allocating resources such as computational power, storage, and network bandwidth across a federated fog system requires sophisticated resource management strategies.

(5) Scalability: As the federation grows, maintaining performance and reliability can become increasingly difficult. Ensuring that the system scales effectively while managing increased complexity is a key challenge.

(6) Latency and Network Overheads: Federated fog environments may involve multiple network hops, potentially increasing latency and network overheads. Optimizing communication and minimizing delays is important for maintaining system performance.

(7) Policy and Governance: Different fog environments may have their own policies and governance models. Harmonizing these policies to ensure smooth operation and compliance across federated nodes can be complex.

(8) Fault Tolerance and Reliability: Ensuring that the federated system is resilient to failures and can recover gracefully from node or network outages requires robust fault-tolerance mechanisms and redundancy strategies.

(9) Management Complexity: Managing a federated fog system involves overseeing numerous nodes, resources, and interactions, which can be complex and require sophisticated management tools and processes.

(10) Compliance and Legal Issues: Federated fog environments may span different jurisdictions with varying regulations and compliance requirements. Navigating these legal and regulatory challenges can be difficult.

Addressing these challenges requires a comprehensive approach that involves advanced technologies, strategic planning, and effective coordination among all participating nodes in the federated fog system.

**VII) Conclusion**

In this survey, we thoroughly investigated the field of fog federation, classifying existing research into three primary areas: federation formation, federation management, and security. Our analysis provided a comprehensive overview of the current state of research, highlighting key contributions and methodologies within each category.We also reviewed existing simulation tools used for evaluating fog federation systems, emphasizing their significance in assessing performance and validating various approaches. Additionally, we identified critical challenges that researchers face today, including issues of interoperability, scalability, and security.By synthesizing these insights, this paper serves as a foundational resource for scholars and practitioners in the field. We encourage future research to address the highlighted challenges and explore innovative solutions, ultimately advancing the capabilities and applications of fog federation in distributed computing environments.

**Disclosure Statement**

The authors have no relevant financial or non-financial interests to disclose.